# Autonomous AI and Ownership Rules

Frank Fagan*

Abstract

As artificial intelligence (AI) systems become increasingly autonomous, traditional notions of ownership will adapt. Historically, property rights have been grounded in traceability, enabling legal and economic systems to allocate ownership efficiently through doctrines such as accession (ownership by connection), and first possession (ownership by labor). Property law also recognizes that abandoned property, when considered untraceable to its original owner, can be efficiently reassigned through accession, ensuring that resources do not remain ownerless. However, autonomous AI systems capable of self-replication, self-governance, and independent economic activity complicate these established principles, raising new questions about how ownership should be determined when AI is no longer clearly linked to an identifiable owner.

This Article examines the circumstances in which AI-generated outputs remain linked to their creators and the points at which they lose that connection, whether through accident, deliberate design, or emergent behavior. In cases where AI is traceable to an originator, accession provides an efficient means of assigning ownership, preserving investment incentives while maintaining accountability. When AI becomes untraceable—whether through carelessness, deliberate obfuscation, or emergent behavior—first possession rules can encourage reallocation to new custodians who are incentivized to integrate AI into productive use. The analysis further explores strategic ownership dissolution, in which autonomous AI is intentionally designed to evade attribution, creating opportunities for tax arbitrage and regulatory avoidance. To counteract these inefficiencies, bounty systems, private incentives, and government subsidies are proposed as mechanisms to encourage AI capture and prevent ownerless AI from distorting markets.

Ultimately, the erosion of traditional ownership structures due to autonomous AI is not merely a theoretical concern but an emerging economic reality. As AI-driven automation expands, ownership may become increasingly provisional, assigned dynamically through legal and economic frameworks designed to maximize efficiency rather than as a default right. This Article argues for an adaptive approach, which would ensure that AI remains within assignable governance structures—whether through accession, first possession, or incentive-driven competition—to prevent the unchecked proliferation of unregulated, unowned AI systems.

What presently [is traceable] to no one becomes by [economic reason] the property of the first taker. — Gaius (paraphrased)[1]

Table of Contents

I. Introduction

Machines have become increasingly autonomous throughout history. During the early stages of the Industrial Revolution, they were used to alleviate the burden of heavy labor.[2] Humans developed machines that leveraged water and steam to supplement hand-production methods.[3] Over time, machines began to take on increasingly complex tasks. Today, fully automated production of subcomponents has become standard practice, while fully automated factory systems are being explored.[4] Moreover, automation has expanded beyond physical labor to include intellectual tasks.[5] Advances in mechanized production during the 20th century set the stage for the mass production of computer chips.[6] Widespread availability, coupled with advances in chip design, brought about an Information Revolution that further expanded the capabilities of machines.[7] Machines moved beyond physical labor to intellectual tasks decisively, including complex information processing and support of human decision-making.[8]

---

1. 1 The Digest of Justinian, bk. 41.1, para 3 (Theodor Mommsen & Paul Krueger eds., Alan Watson trans., Univ. of Pa. Press 1985). The original quote is "[w]hat presently belongs to no one becomes by natural reason the property of the first taker." *Id.*

2. *See* William Rosen, The Most Powerful Idea in the World: A Story of Steam, Industry, and Invention, at xiii (2012).

3. *Id.* at 271.

4. *See* Daniel Kuepper et al., *Can AI Deliver Fully Automated Factories?*, Harv. Bus. Rev. (Aug. 21, 2024), https://tinyurl.com/mr2tz5xn [https://perma.cc/EPX6-UJTM].

5. *See* Stuart J. Russell & Peter Norvig, Artificial Intelligence: A Modern Approach *passim* (3d ed. 2010).

Early forms of automation of information-based tasks required human guidance and supervision, supplementing rather than replacing human cognition.[9] But the latest wave of advancement has brought about systems that no longer merely assist humans, but act autonomously,[10] often in unpredictable and sophisticated ways. Generative models like ChatGPT and Claude, for example, now carry out technical research and writing, computer coding, and even drug discovery.[11] These new capabilities have diminished the role of direct human oversight at a remarkable pace, revealing a continued trajectory in which machines take on an ever-widening range of activities once thought to require complementary human intelligence and intervention.[12]

---

6. *See* PAUL E. CERUZZI, A HISTORY OF MODERN COMPUTING 177 (2d ed. 2003).

7. *Id.* at 346–50.

8. Development enhanced computing power and widespread availability increased the amount of data available for processing. Both advances led to an increase in the capabilities of computing machines. *See* DENNIS YI TENEN, LITERARY THEORY FOR ROBOTS: HOW COMPUTERS LEARNED TO WRITE 2 (2024) (documenting the progression).

9. *See, e.g.*, Frank Fagan & Saul Levmore, *The Impact of Artificial Intelligence on Rules, Standards, and Judicial Discretion*, 93 S. CAL. L. REV. 1, 1 (2019) (providing an example of human-machine collaboration in judging); David A. Hoffman & Yonathan Arbel, *Generative Interpretation*, 99 N.Y.U. L. REV. 451, 451 (2024) (same, with a specific application to estimating contractual meaning with language models); Erik Brynjolfsson, Danielle Li & Lindsey R. Raymond, *Generative AI at Work* (Nat'l Bureau of Econ. Rsch., Working Paper No. 31161, 2023) (documenting the use of AI customer assistants by human call-center agents).

10. *See*, *e.g.*, Guru Angisetty, *How to Use AI to Automate Tasks (+5 Platforms)*, ODIN AI (Nov. 26, 2024), https://tinyurl.com/53cbcy8f [https://perma.cc/4DWU-7RF4] (describing how to deploy a natural language processing system to make decisions independently).

11. On legal research and writing, see generally Frank Fagan, *A View of How Language Models Will Transform Law*, TENN. L. REV. (forthcoming 2025). On coding, see generally Priyan Vaithilingam, Tianyi Zhang & Elena L. Glassman, *Expectation vs. Experience: Evaluating the Usability of Code Generation Tools Powered by Large Language Models*, *in* CHI EA '22: EXTENDED ABSTRACTS OF THE 2022 CHI CONFERENCE ON HUMAN FACTORS IN COMPUTING SYSTEMS (Simone Barbosa et al., eds. 2022). On drug discovery, see Gemma Turon et al., *First Fully-Automated AI/ML Virtual Screening Cascade Implemented at a Drug Discovery Centre in Africa*, 14 NATURE COMM. 5763, 5763 (2023).

12. *See*, *e.g.*, Jan Hatzius et al., *The Potentially Large Effects of Artificial Intelligence on Economic Growth (Briggs/Kodnani)*, GOLDMAN SACHS (Mar. 26, 2023, at 09:05 ET), https://tinyurl.com/mrx3mbzp [https://perma.cc/F4BP-9V53] (noting that AI could expose 300 million jobs, across a wide range of industries, to automation).

The growing autonomy of machines is poised to reshape commercial activity. For an increasing number of goods and services, it is now possible to fully automate both production and distribution processes.[13] Consider, for example, a basic newsletter: machines can automate its creation by scraping the web for articles and organizing them into an email based on pre-programmed instructions. Machines can then automatically distribute the newsletter to email subscribers. When paired with automated payment systems and smart contracts, the possibility of creating a fully autonomous system that operates indefinitely—without any human oversight—becomes a reality. No insurmountable barriers to automated production and distribution stand in the way of continual delivery of the newsletter. The growing potential for perpetual, autonomous provision of services, and even goods, raises important questions about the future of property and the evolving relationship between owners and their agents.

Corporate law empowers and encourages the owners of an enterprise to delegate control over its business operations to managers who are responsible for making key production and distribution decisions.[14] Delegation is valuable because it allows managers the flexibility to adapt and make changes when unexpected events occur.[15] Managers can seize newly presented opportunities and avoid costly mistakes that could arise if decision-making were rigid and pre-programmed. For example, a manager instructed to purchase raw materials at a specific price would miss an opportunity to reduce the company's costs if market prices were to fall. The manager is therefore given decision-making flexibility to purchase materials. While it is straightforward to direct the manager with a rule to purchase at the lowest price, perhaps by a specified date or given contingency, thousands of other factors can disrupt the perfectly timed purchase decision and result in a loss. A supplier, for instance, may pose risks related to the quality and durability of materials, or there may be rumors about the supplier's poor relations with its workforce or regulators. By granting discretion to the manager, the company benefits from flexible decisions that account for millions of previously unknown variables beyond price alone.

---

13. *See id.*

14. Frank H. Easterbrook & Daniel R. Fischel, The Economic Structure of Corporate Law 2–3 (1991) (cataloging the choices corporate managers may exercise dominion).

15. *See* Finn E. Kydland & Edward C. Prescott, *Rules Rather Than Discretion: The Inconsistency of Optimal Plans*, 85 J. Pol. Econ. 473, 473 (1977) (noting that discretion is optimal when policy implementations themselves change the policy environment, including people's behavior).

However, when the parameters of a decision are computable, delegation to managers becomes less valuable.[16] Corporate owners can calculate the optimal decision themselves or instruct a machine to do so continuously, allowing the corporation to operate under rigid and contingent rules applied by the machine. But computability brings its own challenges, and there is a reason no corporation today eliminates managerial discretion.[17] Circumstances change, and some changes are unforeseeable or not fully understood in advance.[18] As long as this remains true, machines will be unable to effectively compute the full decision-space of a corporation and its activities.[19] Nonetheless, as the newsletter example illustrates, opportunities for full or at least adequate computation are increasingly emerging. As automation technology advances, sub-routines that do not require human managers can be spun off into separate entities and governed by detailed instructions contained within a corporation's charter, a corporation's by-laws, or a contract. While it may seem intuitive to assume that a human manager would continue to oversee these sub-routines, albeit from a greater distance, there is reason to believe that corporate owners could delegate this type of big-picture monitoring to machines.[20] The key point is that in certain domains, full automation without any human oversight may eventually become both efficient and commonplace. Acceptance of this premise forms the foundation of the analysis presented here.

At this extreme end of automation, the fundamental fiduciary duties of loyalty and care, which govern the discretionary decisions of corporate managers,[21] become less significant, and the role of law in facilitating the separation of ownership and control diminishes. In an increasingly automated world, a corporate manager begins to resemble more of a trustee than a director. While some level of discretion remains in the protection and preservation of corporate property, traditional corporate decisions rooted in business judgment take on a reduced importance. There is a shift in the law's emphasis from regulating discretionary judgment with corporate law principles to policing protective efforts with the law of trusts.

---

16. *Id.*

17. However, some companies are experimenting with using AI to observe board behavior. *See* Ryan Heath, *AI Shakes Up Corporate Boards*, Axios (Apr. 23, 2024), https://tinyurl.com/22mfyjrf [https://perma.cc/6ET8-KYR6] (describing the use of observational AI on the board of International Holding Company, a valuable United Arab Emirates entity).

18. *See* Frank H. Knight, Risk, Uncertainty and Profit 313–14 (Dover 2006) (noting that if change were subject to unchanging laws, then the future would be computable, and that humans invent the refuge of permutations and combinations to obscure the fact that laws of change can, in fact, change).

19. *See* Frank Fagan, *The Un-Modeled World: Law and the Limits of Machine Learning* 4 MIT Computational L. Rep. 3, 12–14 (2022) (discussing various methods to address incomplete computation and take rules-based action).

20. *See* Heath, *supra* note 17.

21. *Cf.* Theresa A. Gabaldon & Christopher L. Sagers, Business Organizations 501 (3d ed. 2023) (explaining that "all agents of a corporation, in addition to its officers, are fiduciaries and owe duties of care and loyalty").

The impact of growing automation on the laws of business associations and trusts could be substantial and has begun to be addressed elsewhere.[22] In this Article, I would like to focus on the (potential) shock to ownership itself and the problem of its traceability. Consider once more the automated newsletter, now operating without managers. If its record of ownership were to be lost, there is no reason that the algorithm would cease to continue producing and distributing the newsletter, executing payments, and fulfilling its smart contracts. After all, the algorithm is perpetual and self-directed. The only difference would be that the identity of its owners would now be unknown. How should the activities of this class of algorithms be taxed?

Today, this is not a significant problem because owners typically identify themselves and take profits. However, in the future, this may change. Anonymous payment and contracting systems could allow for untraceable profit-taking and create opportunities for tax arbitrage. This is because the income generated by hidden and undiscoverable assets cannot be taxed as easily.

For example, consider an algorithm that offers basic tax filing services for individuals with income below a certain threshold, like those provided by H&R Block. If its ownership becomes obscured, then the filing services could potentially go untaxed. As a result, the algorithm might gain a competitive cost advantage. Over time, this could lead to a decline in filing services that are traceable to an owner, particularly if the owned and unowned algorithms provide identical services or if consumers are otherwise indifferent between using the two. In this instance, the government may step in and offer its own service, as it does today for filers with Adjusted Gross Income of $79,000 or less.[23] As this Article will detail, the provision of a service by the government is a prototypical form of (private) ownership dissolution in response to low-cost automation.

---

22. *See* Noam Kolt, *Governing AI Agents*, 101 Notre Dame L. Rev. 1 (forthcoming 2025) (noting that automated AI agents raise problems of discretionary authority and loyalty that suggest a need for new legal rules); Katja Langenbucher, *Ownership and Trust: A Corporate Law Framework for Board Decision-making in the Age of AI* (ECGI Working Paper No. 758/2024), https://tinyurl.com/yh8p7sz4 [https://perma.cc/B6UH-VX6T] (suggesting that different levels of judicial scrutiny should be applied to AI decision-making deployed by corporate boards); Michael R. Siebecker, *Reconceiving Corporate Rights and Regulation in the AI Era* (Univ. Denv. Legal Stud. Rsch. Paper No. 24-25, 2024), https://tinyurl.com/3r7s2vd3 [https://perma.cc/AJ4D-8U87].

Tyler Cowen has presented another scenario similar to this untraceability problem.[24] Cowen observes that autonomous AI agents[25] are capable of spawning new autonomous AI agents without being directed to do so by a human.[26] This circumstance also creates the potential for tax arbitrage and raises novel questions. Who owns the spawned agents and how should any income that they earn be apportioned? Cowen suggests that no one owns them because they emerge without explicit instructions to come into being by their original owners.[27] If so, then the link between ownership and the provision of services (in this case, bot labor) is severed. Coupled with a payment system and smart contracting,[28] the bot would be able to receive payments, pay server and API[29] access fees, and invest in the development of additional capabilities. Letting autonomous AI labor go untaxed disadvantages human and taxed AI labor and will encourage the proliferation of unowned AI. Cowen thinks this is a bad idea.[30]

This Article takes an alternative view and suggests several approaches to assigning ownership in order to suppress tax arbitrage and other socially costly behavior. It concludes that ownership can be assigned efficiently by relying on the standard tenets of property law. When ownership of autonomous bots is untraceable, then rules of first possession can be applied to encourage new owners to take control.[31] If the costs of acquiring ownership exceed its benefits, then the government or private charities can deploy a bounty system that awards prizes to first possessors.[32]

---

23. *IRS Free File: Do Your Taxes for Free*, IRS (Sep. 15, 2025), https://tinyurl.com/mtwd5pnd [https://perma.cc/DHY7-L7CY].

24. *See* Tyler Cowen, *The Taxman Will Eventually Come for AI, Too*, BLOOMBERG (Apr. 17, 2023, at 07:00 EDT), https://tinyurl.com/398xv7sx [https://perma.cc/2AEE-QVBX].

25. *Id.* Think of agents as bots or algorithms that serve a purpose and are controlled with very general instructions. The discussion to this point has treated these tools as assets. By treating them as agents, Cowen suggests that bots can be analogized to humans or other legal entities. *Id.*

26. *Id.*

27. *See id.*

28. *See* Shawn Bayern, *Of Bitcoins, Independently Wealthy Software, and the Zero-Member LLC*, 108 NW. U. L. REV. ONLINE 257, 263–65 (2014) (noting that banks, and the legal system more broadly, could permit accounts to transact amongst each other so long as knowledge of a passcode is demonstrated; there is no functional need to require that an account bear the title of a person or legal entity).

29. API stands for Application Programming Interface, which is a software intermediary that facilitates communication between two programs. They are often governed by access fees that are incurred on a usage or subscription basis. *See Do APIs Cost Money? Why (and How) Companies Charge for API Access*, OYOVA (July 13, 2023), https://tinyurl.com/mwkhzcu5 [https://perma.cc/8M4T-8XW9].

30. *See* Cowen, *supra* note 24 ("Why own an AI and pay taxes when you can program it to do your bidding, renounce ownership, and enjoy its services tax-free?").

Of course, the availability of any arbitrage opportunity depends on whether autonomous algorithms can do useful work. As alluded to earlier, their number is growing, built largely upon the foundation of large language models.[33] These "agents," sometimes called AutoGPTs, can carry out sales prospecting, perform product and marketing research, develop a business plan, write code, and execute python scripts with little more than very general guidance.[34] If, for instance, a user asks ChatGPT for help planning a birthday party, the language model will respond with a list of things that the user should consider. By contrast, an AutoGPT addresses each aspect of birthday party planning with no need for additional human input.[35] It can develop a theme, guest list, and shop for gifts based on the user's general prompt to "plan a birthday party." These autonomous algorithms that perform individual tasks or collections of tasks present identical questions to the fully automated provision of a good or service discussed earlier: namely, should ownership or some other form of custody be assigned over them—and if yes, how?

This Article explores these questions as follows. Part II examines the emergence of perpetual, autonomous algorithms and autonomous tools more broadly. Part III evaluates the merits of treating autonomous AI as property and considers ownership assignment when the AI is traceable to a creator. The analysis suggests that applying accession doctrine is generally the most efficient approach. Part IV addresses untraceable autonomous AI. When the underlying software resides on a physical server used exclusively by an individual or entity, ownership assignment through accession is efficient. However, if the code exists in the cloud—on servers owned by a large provider of hosting services such as Microsoft Azure, for instance— first possession rules are generally more effective. In such cases, ownership can be granted to the first party to gain decrypted access and take custody of the AI. This approach ensures that investments in improvements and payment of tax obligations occur as swiftly as possible. Part V also explores responses to asocial autonomous AI that imposes social costs, proposing a bounty system as a potential remedy. Part VI concludes.

---

31. *See infra* Part V.
32. *See infra* Part V.
33. *See* Matic, *Auto-GPT vs ChatGPT: How Do They Differ and Everything You Need to Know*, AutoGPT (May 19, 2025), https://tinyurl.com/5aac7bbw [https://perma.cc/2VLH-3DQZ].
34. *See* Nathan Lands (@NathanLands), X (Apr. 12, 2023, at 06:41 ET), https://tinyurl.com/ytfed4tk [https://perma.cc/WQE4-ADM8] (providing examples of AutoGPTs).
35. *See* Matic, *supra* note 33.

## II. Autonomous Tools

For the purposes of this Article, an autonomous tool is defined as a device that performs a useful task without oversight and originates from human conception. Consider a honeybee in the wild. Without supervision, it can pollinate flowers, produce honey, and perform other useful functions vital for the well-being of its ecosystem. However, classifying the honeybee as an autonomous tool is overly broad. Although its origin involves a complex chain of events, it is certain that it was not created by humans. Consequently, it falls outside the scope of this analysis. Once a honeybee is directed to work by a human, such as through honeybee farming, it becomes a tool. But note that farming honeybees requires oversight and thus fails to meet our definition of an autonomous tool. As noted in the Introduction, Cowen observes that some algorithmic tools can be created, or "spawned," by older tools. However, it is crucial to recognize that the chain of their creation can always be traced back to humans. Older algorithmic tools, and certainly the oldest ones, are always developed and produced by human programmers, as algorithms do not spontaneously appear.[36] Throughout this Article, autonomous tools will always be understood to possess the crucial characteristic of having originated from human hands.

It should be noted that the text will sometimes use the terms algorithm, tool, and machine interchangeably. It considers algorithms and machines as species of the genus "tool." Additionally, the text will sometimes use the term automated in place of autonomous, but the reader should always keep in mind the three necessary ingredients of our definition of autonomous tools: usefulness, human origination, and lack of oversight.

---

36. The reader should be comfortable making this assumption given that this form of spontaneity has never been observed. I recognize that the idea of human origination can be weakened on the basis of proximity. Say, for instance, Corporation A creates an algorithm in 2025, which spawns an algorithm one hundred years later. As we will see, welfare is enhanced by assigning ownership, and insofar as A continues to exist, it makes sense to assign ownership to A on the basis of accession. *See infra* Part III. If the new algorithm is untraceable to A, say, for instance, because A has been dissolved, then the algorithm exists in the commons, and a first possession rule maximizes welfare. *See infra* Part IV.

*A. Perpetual Automation*

Most autonomous tools carry out work of limited duration and stop when they finish a task. For example, a translator that instructs a computer to translate a book from French to English sets the computer to work for a limited time. Once the book is translated, the computer automatically stops doing translation work. The job is said to be automated, or pre-programmed, because the human translator does not carry out standard translation tasks such as reading, transcribing into a second language, consulting a dictionary, and so on. This automated work is finite and limited insofar as the translator's instructions are directed to a single book or collection of texts. Once the texts are translated, the algorithm stops until redirected to complete another task. The algorithm's work is not perpetual.

In other settings, work ends when the machine's environment changes. A stock trader that directs an algorithm to buy and sell a stock under a given set of market conditions also directs the computer to work for a limited time. Once those conditions fundamentally change, the algorithm finishes its work and stands by for future instructions. In yet other circumstances, an automated machine can be directed to carry out a task indefinitely so long as it is able. A rover on Mars or a standard space probe will send exploration data back to Earth on the condition that it retains sufficient energy and operability. When the machine ceases to function, it no longer carries out work. Thus, automated tasks are finite in duration because we want them to be or because the machines that execute them (or their energy sources) are exhaustible. Today, the limited duration of automated work is standard and commonplace.

In the future, automation will be more open-ended and indefinite. Reconsider the automated newsletter that aggregates news stories. The algorithm scrapes stories of current events from the Internet, organizes them by topic, and sends them to paying subscribers in the form of an email newsletter. Once programmed, all these tasks can be completed without human intervention. The owner of the algorithm simply manages the mailing list and collects the fee. While this example of digital publishing may appear to showcase novelty in automation, it shares a lot in common with the physical vending machine conceived decades ago. Consider that once the distributor of the online newsletter accumulates some subscribers, she can earn a profit without making any additional effort to gain new customers. Similarly, once the vendor has installed his machine at some location and filled it with goods, he simply waits for passersby to transact with the machine. In both examples, distribution is automated insofar as neither requires a human to interface with a customer. The distribution process is programmed and then left alone. Along the same lines, the process of new customer acquisition can be automated. The newsletter manager might post an enrollment form online, for instance, that is able to sign up new customers. This method is identical to a vendor who places a machine in some outpost, say, a subway platform, where his customers can transact. The task of customer acquisition is automated. The broader point is that a substantial portion of a product's distribution chain, if not all of it, can be automated from start to finish.

The significant difference between the two examples is not with their methods of provisioning and distribution, but rather with their approach to production. The manager of the newsletter produces her product through automated scraping of the Internet. The algorithm sorts the stories into topics and prepares an email for bulk transmission. At no point does the manager need to engage in labor associated with the newsletter's creation. By contrast, the vendor must occasionally complete the work of filling an empty machine with food and drink. While both automate distribution, the newsletter fully automates production. Nowadays there is nothing extraordinary about automated production. Industrial automation has been growing since the emergence of the machine, and the arrival of information goods in the last several decades has brought many opportunities for merchants to automate entire production processes from start to finish.[37] What is presently remarkable is that machines are rapidly encroaching on many forms of human labor across information- and knowledge-based industries such as computer programming, teaching, customer service (via chatbots and virtual assistants), mechanized data entry and analysis, algorithmic financial trading, medical imaging analysis, drug discovery, content recommendation (especially entertainment content), and many others.[38] This broad encroachment of machines into information-based industries will bring new opportunities for automation of entire distribution and production chains.

When these chains are fully automated, human discretion can be handed over to the machine. Consider again the newsletter. Its production can be periodically fine-tuned to account for changes in customer tastes, perhaps provoked by observing which news articles are viewed the longest by its readers or by keeping track of which types of stories are shared with others more often. The human manager no longer acts on the basis of customer reactions, the business environment, and other factors. No Netflix employee adjusts recommendations based on customer habits.[39] The Netflix algorithm, which can rely upon a simple system of like-or-dislike input from the user, is given the power to select what the customer sees.[40] In this way, production can be adjusted to customer tastes and needs without human decision-making or oversight. Distributional flexibility can be automated just as well. Prices, for instance, can be indexed to inflation, product demand, and the competitive environment more generally.[41]

---

37. On the history of industrial automation, see Barry C. Brusso, *Fifty Years of Industrial Automation*, 24 IEEE INDUS. APPLICATIONS MAG., June 8, 2018, at 8–11. On the automation of entire production processes, see Kuepper et al., *supra* note 4.

38. *See* Jan Hatzius et al., *supra* note 12; *see also* Bayern, *supra* note 28 at 263 (providing additional examples of autonomous computer tasks).

---

39. *See How Netflix's Recommendations System Works*, Netflix, https://tinyurl.com/yfnxavyk [https://perma.cc/WZQ3-D9MD] (last visited Feb. 8, 2025) ("[W]e take feedback from every visit to the Netflix service . . . and continually update our algorithms with those signals to improve the accuracy of their prediction of what you're most likely to watch.").

40. *See id.*

41. Tax law has deployed this type of flexibility for years through inflation indexing. *See* Daniel Shaviro, *The More It Changes, The More It Stays the Same?: Automatic Indexing and Current Policy*, *in* The Timing of Lawmaking 64, 64 (Frank Fagan & Saul Levmore eds., 2017). There are other examples in law. Government assistance is made contingent on income and age; tax credits are made contingent on desirable behaviors; and penalties are levied on the undesirable. For an effort to make areas of law other than tax more contingent, see generally Frank Fagan, *Legal Cycles and Stabilization Rules in* The Timing of Lawmaking 11 (Frank Fagan & Saul Levmore eds., 2017).

Where automation faces resistance, it can be overcome through the partitioning of tasks into smaller components, providing greater flexibility and adaptation that assuages customer concerns, and by relying on user input. Consider again the algorithm that provides basic accounting and personal tax preparation for individual filers. Perhaps this example seems a bit far-fetched given that many consumers prefer working with a human in case they are audited. But tax preparation and audit services can be partitioned. One can be given to a machine and the other to a human. The basic approach is to separate the service into components so that the customer is comfortable with the machine doing one part and a human doing another. Similarly, services can be adjusted to provide for flexible and adaptive solutions that include updates. Many already exist. A computer operating system adapts to user behavior, anti-virus software routinely updates its files to account for new threats, and a retirement account responds to its holder growing older by routinely rebalancing a portfolio. These updates may presently require a human to program them, but that work, too, can be automated. For example, the operating system can observe user behavior and compile periodic reports, which are then used to reprogram the software or add new features (without human input). Viruses, too, can be scraped from the Internet like the newsletter articles, and anti-virus software can respond.

Further, many services today require substantial work on the part of the customer. In some instances, an algorithm must learn from its user as when Netflix must be told whether the customer approves or disapproves of a new show. A human must enter data. In other instances, a user must complete more substantial work as when tax preparation software requires detailed input about income and assets. Nonetheless, it is not difficult to imagine the production of a service like tax preparation partitioned between humans and machines with the former farmed out entirely to the customer. From the perspective of the service provider, automation that requires

Users of Uber are familiar with this form of automation. Purchasing a ride during a surge in demand is more costly than when demand is low, and many drivers are available.[42] Uber's surge pricing, and more generally its distribution of ride services, can be said to be automated insofar as its algorithm is self-executing.[43]

## B. Eroding Oversight

When distribution and production chains are fully autonomous, most, if not all, managerial discretion can be transferred to a machine. Liability generated by autonomous tools can be controlled by the usual channels of tort and contract law,[44] as well as corporate law insofar as corporate directors are used to monitor these automated processes. For example, consider a corporation engaged in automated tax preparation services that relies on an algorithm to periodically scan the tax code to adapt its advice to the latest rules. The corporate board or a director could be tasked with determining whether the algorithm is sufficiently accurate to update its knowledge base independently or whether human tax specialists should double-check the algorithm's revisions. This oversight task might be necessary to fulfill a fiduciary obligation.[45] At a minimum, a large corporation might be expected to implement a monitoring program to ensure compliance with the law.[46]

However, if autonomous tools are truly independent and their capabilities exceed those of the human tax specialists, then it seems likely that directors could fulfill their duties without deploying additional human verification. Their monitoring function would simply involve ensuring that the algorithm's capability remains superior to that of humans. This might be accomplished with minimal effort, such as ensuring that the tax software has been updated. As in other contexts, managers should be permitted to lawfully rely on the honesty

---

human input by customers (only) is tantamount to automation of its share of responsibilities. The overarching point remains that complete automation of distribution and production chains is achievable, notwithstanding debates over finer details.

42. *See How Surge Pricing Works*, Uber, https://tinyurl.com/5a2vdrxr [https://perma.cc/2694-QUQG] (last visited Feb. 8, 2025).

43. *See id.* (explaining that, "[b]ecause rates are updated based on the demand in real time, surge [pricing] can change quickly").

44. *See, e.g.*, Leslie Y. Garfield Tenzer, *Defamation in the Age of Artificial Intelligence*, 80 N.Y.U. Ann. Surv. Am. L. 135, 140–41 (2024) (explaining that concerns that generative AI will upend the "tort of defamation" are "unfounded" and arguing that "plaintiffs can bring defamation claims stemming from generative AI without reforming defamation law").

45. *See* Langenbucher, *supra* note 22, at 13–15 (noting that corporate boards which delegate decisions to AI must maintain some discretion to overrule the AI in order to carry out their fiduciary obligations).

46. *See In re* Caremark Int'l, Inc., 698 A.2d 959, 969–70 (Del. Ch. 1996).

and integrity of their subordinates, whether human or algorithmic, so long as they have good faith reasons to remain confident in those subordinates.[47] If we believe that algorithms and automation will continue to advance, it is easy to see that at directorial oversight over certain tasks will eventually become redundant. No one expects an accountant to verify the math of today's pocket calculator or Excel spreadsheet by hand. Corporate boards and other forms of human oversight will remain essential insofar as humans remain superior, or at least competitive, in completing certain tasks, including the task of monitoring itself. However, as machines become capable of performing more tasks better than humans, the need for directorial oversight will erode.

Thus, whether it will be possible to successfully and competitively operate an entity without any human oversight at all is a question of technical feasibility. One might accept the premise of eventual feasibility on the basis of the rapid technological advance that has taken place within the field of artificial intelligence over the past decade.[48] In a legal and economic system such as ours, entrepreneurs and innovators are encouraged to continually develop sophisticated machines that can outperform humans in all areas of life.[49] From this perspective, the erosion of oversight is just a matter of time.

Another way to accept the feasibility premise is by conceding that an ability to overcome the challenge of discretionary judgment is just a matter of degree. Rules-based decision-making devoid of discretion may be possible for managing some businesses, but not all. And parts of a business that are amenable to being managed by rules—and governed by a duty of obedience[50]—can be spun off and made into separate entities.

---

47. *See* Graham v. Allis-Chalmers Mfg. Co., 182 A.2d 328, 332 (Del. Ch. 1962).

48. For a similar argument made for the eventual dominance of machines in law because of the computability of finely tuned rules, see Anthony J. Casey & Anthony Niblett, *The Death of Rules and Standards*, 92 Ind. L.J. 1401, 1401 (2017).

49. Some commentators have even asserted that societies should push law (in some areas) toward removing humans altogether from "the loop" of decision-making because human oversight is thought to create more problems than benefits. *See* Orly Lobel, *Automation Rights: How to Rationally Design Humans-Out-Of-The-Loop Law*, Univ. Chi. L. Rev. Online (2024), https://tinyurl.com/2t333vzu [https://perma.cc/ZAC3-ZJF9]. The point is that areas where human discretion was once considered a necessary ingredient may give way to rules-based architectures that can serve as effective substitutes. *Compare* Charlotte A. Tschider, *Humans Outside the Loop*, 26 Yale J.L. & Tech. 324, 324 (2024) (noting that humans will remain in the loop despite efforts to remove them) *with* Commission Regulation 2016/679 of Apr. 27, 2016, General Data Protection Regulation, art. 21, 2016 O.J. (L 119) (EU) (providing for a right to human oversight of a machine decision).

50. For a discussion of this often-overlooked duty and an attempt to revive it, see Alan R. Palmiter, *Duty of Obedience: The Forgotten Duty*, 55 N.Y. L. Sch. L. Rev. 457, 457 (2011).

III. Traceable Ownership

Traceable entities without managers, whether they complete entire tasks from start to finish or cabined sub-routines of production and distribution tasks, seem unlikely to complicate existing patterns of accountability for at least two reasons. First, permitting entities to exist without human managers pushes the boundaries of existing law.[51] Such entities are likely prohibited. If so, then there is no need to develop methods for taxing and holding them liable for their acts. Responsibility remains in the hands of human managers and owners. Second, the spun-off entities in charge of sub-routines, even if granted separate legal status, can continue to be tethered to their creators through existing rules such as those governing successor liability, or even veil-piercing, so long as a creator possesses some connection to the manager-less entity.[52] Thus, existing rules for assigning income and liability are readily available to successors and subsidiaries, even if law permits these to be governed without human managers. A more difficult challenge arises when ownership is not traceable.

---

51. Bayern, *supra* note 28, notes that zero-member LLCs can arise in common circumstances that leave gaps in membership and that the Uniform Limited Liability Company Act (ULLCA) recognizes this possibility explicitly. *Id.* at 267–68. The Act provides for automatic dissolution when an LLC has no members for 90 consecutive days. *See* Unif. Ltd. Liab. Co. Act § 701(3) (2013) (Unif. L. Comm'n). However, automatic dissolution in this circumstance is not mandatory.
A corporation must be managed by a board of directors, and all directors must be natural persons. In Delaware, for example, "[t]he board of directors of a corporation shall consists of one more members, each of whom shall be a natural person." 8 Del. Code Ann. § 141(b). This limitation exists across American jurisdictions. *See, e.g.*, Model Bus. Corp. Act § 8.03(a) (2002) ("A board of directors must consist of one or more individuals . . . ."); *id.* § 1.40(13) ("Individual means a natural person."). While the Model Business Corporation Act permits a group of unanimous shareholders to eliminate the board, the shareholders still retain control over the corporation's actions. *Id.* § 7.32(a). Thus, even without a board, ultimate corporate authority remains with legal persons. Legal persons can consist of other entities of course, severing a direct link between a human shareholder-manger and the director-less entity. But a second, indirect link would exist between the director-less entity and the human managers who control the purchaser of its shares. *See* Shawn Bayern, *The Implications of Modern Business-Entity Law for the Regulation of Autonomous Systems*, 19 Stan. Tech. L. Rev. 93, 99–100 (2013) [hereinafter *The Implications of Modern Business-Entity Law*] (noting that autonomous systems serving as corporate legal persons would require "the ongoing consent [and presumably oversight] of an existing private party").
Partnership law provides a possible avenue to manager-less entities. Two natural persons can first form a general partnership, then enter into an operating agreement that requires the partnership to carry out the actions dictated by an autonomous tool, and then finally dissociate from the partnership. RUPA notes that "[t]he association of two or more persons to carry on as co-owners a business for profit forms a partnership." Unif. Partnership Act § 202(a) (1997) (Unif. L. Comm'n) [hereinafter "RUPA"]. The operating agreement governs their relations and can contain anything except blacklisted terms contained with section 103(b)(8). *See id.* § 103(a). Governance by an autonomous tool is not listed. *Id.* § 103(b)(8); *see also The Implications of Modern Business-Entity Law*, *supra*, at 101. However, it is unclear whether the partnership without human partners would be able to lawfully continue, though it is perhaps unlikely, since partnerships have historically consisted of associations of two or more persons. For discussion, see *id.* at 100–01. *See also* Robert W. Hillman & Donald J. Weidner, *Partners Without Partners: The Legal Status of Single Person Partnerships*, 17 Fordham J. Corp. & Fin. L. 449 (2012) (discussing whether partnerships with one partner can persist under RUPA).
A surer path to a manager-less entity is through the more flexible Limited Liability Company (LLC). As before, a natural person can first form a member-managed LLC, then enter into an operating agreement that requires the LLC to take actions specified by an autonomous tool, and then finally withdraw membership. As noted in *The Implications of Modern Business-Entity Law*, *supra*, at 101 n.30, the assignment of governance to an autonomous tool, followed by withdraw, can be accomplished with either a member-managed or manager-managed LLC.

Before turning to the challenges presented by untraceable ownership, consider in detail the spawning scenario raised by Tyler Cowen. Tools, such as self-executing computer programs, can create autonomous tools without being directed to do so by a human or human-managed entities. Cowen suggests that the tools intentionally created by humans are their property, but the unintentionally created ones are not.[53] These spawned tools could be considered independent and

---

There is reason to think that an LLC without managers, as opposed to a partnership without partners, is more likely to be permitted under existing law. The Uniform Limited Liability Company Act (ULLCA) permits an LLC with no members to continue operating for 90 days. *See* Unif. Ltd. Liab. Co. Act § 701(a)(3) (2013) (Unif. L. Comm'n) (noting that court-ordered). Furthermore, ULLCA Section 105(c)(9), which enumerates the statute's non-waivable means of dissolution, does *not* include the 90-day limitation on member-less LLCs despite mentioning other scenarios that trigger mandatory dissolution. Specifically, section 105(c)(9) notes that the operating agreement cannot vary the terms of dissolution set out in section 701(a)(4). The mandatory grounds for dissolution are fraud, illegality, and oppression. *Id.* Students of statutory interpretation will recognize a form of *expression unius* at work. If the legislature wished to explicitly prohibit zero-member LLCs, then it would have provided for mandatory dissolution following the permitted 90-day window. The fact that the statute expresses grounds for dissolution as fraud, oppression, or general illegality, but not for having no members, suggests that the legislature had no intent to include managerial exit as a reason for dissolution. *See* William N. Eskridge, Jr., Philip P. Frickey & Elizabeth Garrett, Cases and Materials on Legislation, Statutes, and the Creation of Public Policy 854–55 (4th ed. 2007) (describing methods of interpreting implied intent). Shawn Bayern notes that New York has come close to making explicit the possibility of creating perpetual manager-less entities. Its statute permits a zero-member LLC to continue for 180 days "or such other period as is provided in the operating agreement after the occurrence of the event that terminated the continued membership of the last remaining member . . . ." *The Implications of Modern Business-Entity Law*, *supra*, at 103 (quoting N.Y. Ltd. Liab. Co. Law § 701(a)(4) (1999)).

52. Predecessor entities can be liable for their successors' acts, especially when successors enable liability evasion. *See* Frank Fagan, *Successor Liability from the Perspective of Big Data*, 9 Va. L. & Bus. Rev. 391, 391 (2015). Similarly, veil-piercing is often invoked to pin liability on an opportunist. *See* Jonathan Macey & Joshua Mitts, *Finding Order in the Morass: The Three Real Justifications for Piercing the Corporate Veil*, 100 Corn. L. Rev. 99, 99 (2014).

53. *See* Cowen, *supra* note 24.

autonomous—and therefore ownerless—even if their lineage is readily apparent and traceable. This is a remarkable proposition given that assigning ownership is generally regarded as socially beneficial.[54]

Consider the spawn scenario in the following context. Suppose a human-controlled corporation, Parent, Inc., creates an autonomous tool, Orphan AI, that analyzes a patent database for gaps in inventions that could link two or more existing inventions together. After discovering a gap, Orphan AI drafts a new interfacing patent, files it with the patent office on behalf of Parent, and then markets licensing arrangements to businesses that own the unlinked patents. Today, only natural persons can be inventors.[55] For the purposes of the hypothetical, assume that Orphan AI merely suggests interfacing patents to Parent, Inc. employees, who lawfully process them in turn.[56]

Suppose further that Parent furnishes the interfacing patent tool (Orphan AI) with a general directive to generate profit. Unknown to Parent, the tool has discovered a number of unfiled interfacing invention. In response to Parent's instruction to generate profit, Orphan AI then spawns a new tool to take advantage of this nascent intellectual property. Orphan AI sets up a new entity, Orphan, Inc., and then produces and distributes a medical diagnostic software tool wholly dependent upon the unsold interfacing discoveries. Orphan, Inc. keeps the underlying diagnostic algorithm secret (since it knows as a non-human it cannot lawfully protect it with its own intellectual property rights). The software generates substantial profits. How should law treat Orphan AI and Orphan, Inc.?

Orphan AI is clearly the property of Parent, Inc., but the legal relationship is strained because Orphan possesses some characteristics common to agents. Parent, Inc. provided it with express authority to act on its behalf when it instructed Orphan to file and market the patents. Further, Orphan was initially acting for the benefit of Parent before it transferred the patents to a new entity. At this point in the hypothetical, it may be tempting to treat Orphan as an agent, but people and corporations often use automated tools to complete tasks, and we rarely, if ever, think of these tools as agents that bind a principal. The difference here is that Orphan possesses a greater degree of autonomy than, say, industrial robots that perform automated manufacturing, or automated scripts that send mass-marketing emails. Its enhanced freedom stems from its technological capabilities, powered by its generative AI architecture, as well as the overarching directive from Parent to "generate profit." This directive grants it a level of autonomy commonly associated with agents.

---

54. *See infra* note 92.
55. *See* Thaler v. Vidal, 43 F.4th 1207, 1209 (Fed. Cir. 2022); Matt Blaszczyk, *Impossibility of Artificial Inventors*, 16 Hastings Sci. & Tech. L.J. 73, 75–76 (2024).
56. *See, e.g.*, Inventorship Guidance for AI-Assisted Inventions, 89 Fed. Reg. 10043, 10046 (Feb. 13, 2024) (noting how combined human-machine efforts can lead to recognized inventions); Yuan Hao, *The Rise of "Centaur" Inventors: How Patent Law Should Adapt to the Challenge to Inventorship Doctrine by Human-AI Inventing Synergies*, 104 J. Pat. & Trademark Off. Soc'y 71, 133 (2024) (same).

Nonetheless, it makes little sense to distinguish Orphan as an agent because Parent will undoubtedly require Orphan to transfer any generated profits back to it (perhaps such a directive is implemented in the original code that Parent wrote to create Orphan), erasing any meaningful distinction between property and agent. Whether Orphan is viewed as an economically productive asset or an agent, the outcome remains the same: Parent profits. Furthermore, any liability arising from Orphan's initial actions, prior to the creation of Orphan, Inc., remains the same. If treated as property, Parent is responsible for liabilities arising from its use of Orphan AI as a tool. If considered an agent, all of Orphan's actions (patent gap identification, patent filing, marketing, and licensing), would very likely fall within the scope of Parent's authority since these actions were programmed by Parent.[57] Again, the distinction is without difference.

Once Orphan AI creates the separate Orphan entity and the medical diagnostics software, and begins to take profits for itself, the analysis may seem to change, but it does not. At the outset, Orphan AI cannot establish a corporation without legal persons to oversee its management under existing rules.[58] It might theoretically create an LLC or partnership without members or legal persons. But these entities are temporary, and the extension of their lawfulness beyond a grace period of 90 days remains uncertain.[59] Assuming that existing rules remain the same and do not permit the creation of entities without legal persons, then the creation of a for-profit entity like Orphan, Inc. would be barred and the hypothetical ends here.

Instead, assume (wildly?) that law is modified to permit the creation of entities without legal persons as managers or owners.[60] The AI can now create Orphan, Inc. and take profits. For the moment, set aside the question of who owns this entity. Orphan, Inc. is clearly violating Parent's intellectual property rights, and so long as Parent is aware, Parent can recover the value of Orphan, Inc.'s unauthorized use of Parent's intellectual property rights.[61] All the economic output of Orphan essentially belongs to Parent insofar as the value of the diagnostic software is completely dependent on Parent's patents. Before relaxing this assumption, consider any potential liability for activities arising out of Orphan, Inc. Once Orphan AI creates the entity, and it begins its operations, the hypothetical introduces a break in the chain of Parent's control. This break may not be foreseeable, or avoidable with responsible programming of Orphan AI by Parent. Nonetheless, a strict liability standard could be applied to assign responsibility for Orphan AI's acts to Parent.[62] The point is that liability can be assigned by relying on familiar objective standards found elsewhere in law.[63] The more complicated question is what to do about ownership of Orphan AI's outputs, again, under the assumption that law permits the creation of Orphan, Inc.

---

57. I recognize that there are countervailing opinions, especially if the AI were to clearly act outside the scope of Parent's authority, say, for instance, by ignoring explicit instructions. However, Parent can be held liable for failing to exercise due care in supervising or training Orphan. *See* Ian Ayres & Jack Balkin, *The Law of AI Is the Law of Risky Agents Without Intentions*, U. Chi. L. Rev. Online *1, *2 (2024) ("People may also be held responsible for designing, programming, or training a risky AI program that causes harm to others.").

58. *See supra* note 51. Legal persons can be directors, shareholders, or other entities managed by directors or shareholders.

59. *See supra* note 51.

60. Another (wild) possibility is to grant legal personhood to Orphan AI and allow it to manage Orphan, Inc.

### A. The Laws of Accession

If Parent owns Orphan, Inc., then it clearly owns the outputs. There is a clear doctrinal channel for assigning ownership of Orphan to Parent: accession. Accession is like a magnet.[64] The doctrine of accession awards a prize to a person (or in this case, an entity) who owns another resource prominently connected to that prize. Newborn animals are prominently connected to the dam or mother, crops are so connected to their soil, and new soil is attached to its lowered riverbanks.[65] In each of the aforementioned examples, a meaningful connection is present.

Social norms determine whether a connection is prominent and meaningful. Therefore, social norms can be understood as the core determinant of ownership from the perspective of accession.[66] Ownership of unintentionally spawned AI has not (yet) been disputed. As an initial matter, it is difficult to imagine a dispute in the first place. Once Parent learns of Orphan's outputs, it can assert itself and simply take control of Orphan, Inc., so long as it has access to Orphan's code. As an entity, the spawned AI could, in theory, attempt to defend itself by bringing a conversion claim against Parent. But this seems fanciful if not procedurally impossible.[67] If such a brave new world arrives, then the case will turn on whether Orphan is prominently connected to Parent so that the laws of accession apply. A more plausible scenario in the near future involves instances where Parent is unaware of Orphan's existence or, more broadly, where autonomous tools cannot be traced back to a clear and identifiable source. As we will see, the laws of first possession can be applied to assign ownership.[68] But let us continue this discussion as if the brave new world were imminent.

---

61. Under 35 U.S.C. § 284, courts shall award damages "adequate to compensate for the infringement, but in no event less than a reasonable royalty for the use made of the invention by the infringer." 35 U.S.C. § 284. Courts may also issue injunctions to prevent ongoing or future use. 35 U.S.C. § 283; eBay Inc. v. MercExchange, L.L.C., 547 U.S. 388, 391 (2006) (setting out a four-factor test to determine appropriateness of injunction).

62. *See* Ayres & Balkin, *supra* note 57 at *2 (noting the possibility).

63. *Id.* at *1 ("We think that the best solution [for determining AI liability] is to employ objective standards that are familiar in many different parts of the law.").

64. This distinction is made in Thomas W. Merrill, *Accession and Original Ownership*, 1 J.L. Analysis 459, 463 (2009).

65. *Id.*

66. *See generally* Robert C. Ellickson, Order without Law: How Neighbors Settle Disputes (1991) (describing how social conventions influence ownership rules).

While social norms surrounding artificial intelligence are currently in flux, examining other examples of accession, as well evaluating the economic rationale for assigning ownership to Parent, can be instructive for ascertaining whether it should be permitted to assume possession of Orphan, Inc.

Table I: The Family of Accession Doctrines

| Newborn animals | "The general rule, in the absence of an agreement to the contrary, is that the offspring or increase of tame or domestic animals belongs to the owner of the dam or mother . . . ."[69] |
|---|---|
| Crops | *Fructus naturales* (trees, bushes, and other perennials and their fruits belong to the owner of the soil; *fructus industriales* (annual crops that require planting and harvesting) belong to the owner of the soil[70] |
| Accretion | Land that is gradually augmented by alluvial formations belongs to the riparian landowner[71] |
| Commingled goods | Where there are two owners of inputs that create an output, the good belongs to the owner that has supplied the more valuable input[72] |
| Ad coelum | Underground minerals, caves, and ground water belong to the owner of the surface[73] |
| Fixtures | Chattel that "bears a relationship to the land so physically close or otherwise prominent" belongs to the owner of the land[74] |
| Interest | Uncontracted for interest earned on a fund of money, e.g., escrow funds, security deposits, and prepayments for services, belong to the owner of the principal[75] |

---

67. This question is left for interested civil procedural specialists.
68. *See infra* Part IV.
69. Carruth v. Easterling, 150 So. 2d 852, 855 (Miss. 1963) (quoting 4 Am. Jur. 2d *Animals* § 10).
70. *See* Merrill, *supra* note 64 at 465.
71. Nebraska v. Iowa, 143 U.S. 359, 360 (1982).
72. *See* Merrill, *supra* note 64, at 466.
73. *Id. at* 467.
74. *Id*.
75. *Id.* at 468 (first citing Brown v. Legal Found. of Wash., 538 U.S. 216, 235 (2003); then citing Philips v. Wash. Legal Found., 524 U.S. 156, 164–71 (1998); and then citing Webb's Fabulous Pharmacies v. Beckwith, 449 U.S. 155, 162–64 (1980)).

| | |
|---|---|
| Adaptation rights | Holder of copyright holds exclusive right to create derivative works[76] |
| Patent improvements | Patentee holds exclusive rights to refine invention and block anyone else who uses the invention to create something new; patentee also holds exclusive claims to minor changes to patent under the doctrine of equivalents[77] |
| Trademark dilution | Owner of famous marks protected against dilution so long as a mark has a close enough connection to do a famous mark potential harm[78] |
| Publicity | Celebrities hold exclusive rights to the use of their images in a particular commercial market[79] |

These examples of the principle of accession are well settled. In each instance, law has essentially declared that the connection between the primary and derivative resources is sufficiently prominent. Merrill generalizes further and notes that accession can be understood to allocate new resources as well as increments of value.[80] The owner of a large tank of oil obviously "owns" the windfall generated by a surge in oil prices. The owner of once cheap farmland "owns" the windfall generated by a surge in value as the city expands towards the farm. The connection is so undeniably close that we assign value without giving it much thought. However, when a new resource or increment in value is somewhat attenuated from its source, so much so that we begin to think of it as a separate asset, do we then turn to one of the categorical examples above.

Autonomous AI does not fit neatly into any of the categories above and can be easily distinguished. Newborn animals, for instance, are not fully developed and require the care of their mothers. Autonomous AI is fully independent. Both *fructus naturales* and *fructus industriales* require physical proximity and nutrients from their soil until harvested, and last only for a time before they spoil. Autonomous AI can last perpetually on its own, far away from its original creator. The same reasoning can be used to distinguish fixtures, accreted land, and underground resources. Each of the derivative resources in these examples are physically proximate to the primary resource. Autonomous AI need not be. The examples of intellectual property and celebrity publicity can easily be distinguished, too. In each of them, the derivative resource contains or resembles the primary one. Thus, patent improvements maintain a core invention. Adaptations are recognizable versions of an original. Infringing trademarks or a celebrity's image bear some likeness to a famous mark or person. By contrast, the attributes of autonomous AI can be entirely disjoint and unrelated to the attributes of its creator. Of course, there are similarities between these examples and autonomous AI, but the point is that the latter can be readily distinguished. This indicates that the principle of accession can be flexibly used to establish both the independence as well as ownership of Orphan, Inc.

---

76. 17 U.S.C. § 106(2).
77. Merrill, *supra* note 64, at 468–69 (first citing Robert P. Merges & Richard R. Nelson, *On the Complex Economics of Patent Scope*, 90 COLUM. L. REV. 839, 865–66 (1990); and then citing Mark A. Lemley, *The Economics of Improvement in Intellectual Property Law*, 75 TEX. L. REV. 989, 1070 (1997)).
78. *Id.* at 469 (citing Clarissa Long, *Dilution*, 106 COLUM. L. REV. 1029, 1034 (2006)).
79. *See id.* at 469 (citing Midler v. Ford Motor Co., 849 F.2d 460 (9th Cir. 1988)).
80. *Id.* at 473.

Perhaps the most interesting comparison is to commingled goods. Section 9-336(a) of the Uniform Commercial Code notes that sometimes goods can be "physically united with other goods in such a manner that their identity is lost in a product or mass."[81] While the identity of autonomous AI can be separated from its creator, the rationale for ownership of a commingled good depends on the magnitude and value of the labor contributions of its producers.[82] A more significant and valuable contribution can be seen as establishing a stronger connection. Orphan's contribution is comparable to labor: it searches patent databases, identifies interfacing techniques, secures them as trade secrets, and markets them to third parties. Parent's contribution lies in the creation of Orphan itself. While Orphan's output may surpass Parent's in scope, the value of Parent's contribution appears greater and more fundamental, as Orphan relies entirely on Parent's initial input to function. Rules for commingled goods, therefore, seem to imply that Parent should own Orphan's outputs.

There are several examples that illustrate the principle of accession in transition. These transitions can be understood as a reflection of changing social norms that determine the presence of a prominent connection. Domain names on the Internet, for instance, were initially assigned on a first-come, first-served basis.[83] But cunning opportunists began purchasing domain names associated with prominent trademarks (and tradenames) and then offered them for sale at elevated prices.[84] This practice, known as cybersquatting, was challenged by trademark owners. Courts essentially held that an owner's rights in traditional channels extended to the Internet.[85] Congress ratified these decisions by enacting the Anti-Cybersquatting Protection Act of 1999 (ACPA).[86] Thus, a rule grounded in first possession gave way to rule based upon accession.

---

81. U.C.C. § 9-336(a) (A.L.I. & Unif. L. Comm'n 2010).

82. The Romans addressed the problem of commingled goods with three separate doctrines. Merrill, *supra* note 64, at 466. They applied *accessio* when two different items were joined into one, as when wine owned by *A* was placed in bottles owned by *B*. *Specificatio* applied when raw material such as grapes was transformed into a different product (wine) by *B*'s labor. And they applied *confusio* when two or more persons contributed similar inputs whose contributions could not be discerned, such as when *A*, *B*, and *C* placed grapes into a single vat for fermentation. Merrill notes that American courts apply accession to factual scenarios of *accessio* and *specifcatio*. *Id.* Accession is distinguished from *confusio*. All three of these scenarios, however, can cover situations in "which [the] owner of inputs has supplied the larger or more valuable input." *Id.*

83. *See* Michael Karanicolas, *The New Cybersquatters: The Evolution of Trademark Enforcement in the Domain Name Space*, 30 Fordham Intell. Prop. Media & Ent. L.J. 399, 399 (2020) (noting the history of domain name assignment).

Street parking offers another example. In areas with low congestion and light usage, parking is typically allocated on a first-come, first-served basis.[87] However, as space becomes limited, local communities often introduce permit systems that reserve parking for permit holders. Typically, the right to park is tied to residency in the area, which serves as the "prominent connection." But in both examples—domain names and parking—the prominent connection is always there. Law moves from the first-come, first-served principle to accession because something else has changed—namely, the economic stakes of possession: domain names and parking spaces become salient and more valuable. None of this seems particularly instructive for assigning ownership of Orphan (and its outputs) because the economic stakes for Parent remain constant over time. When traffic on the Internet and street increases, cyber and physical real estate becomes more valuable. In the hypothetical, Orphan is consistently successful from the start. It is true that the value of Orphan's trade secrets could at first be trivial and then increase with Orphan's growing ability to market them. Alternatively, Orphan could benefit from exogenous shocks that increase demand. While this line of reasoning opens the door to rationalizing ownership by Orphan, it cannot sidestep the fact that Parent, as Orphan's creator, is responsible for Orphan's output at bottom. The commingled goods analogy is difficult to overcome.

As noted above, law could assign a comparatively greater value to Orphan's efforts. But to do so, it must acknowledge that the value of creating Orphan is less, even though creation is obviously essential for Orphan's functioning. Of course, a method of comparison other than value could be used just as well. Hume suggests several candidates: "One part of a compound object may become more considerable than another, either because it is more constant and durable; because it is of greater value; because it is more obvious and remarkable; because it is of greater extent; *or because its existence is more separate and independent.*"[88] Note this last justification of separateness and independence could easily justify assigning ownership to an autonomous Orphan. Law must select from opposing rationales. One way to resolve the choice is to select the rationale that maximizes social welfare.

---

84. *See id.*

85. *See* Merrill, *supra* note 64, at 472 (first citing People for the Ethical Treatment of Animals v. Doughney, 263 F.3d 359 (4th Cir. 2001); and then citing TCPIP Holding Co. v. Haar Communications, Inc., 244 F.3d 88 (2d Cir. 2001)).

86. 15 U.S.C. § 1125(d); *see also* Prudential Ins. Co. of Am. v. Shenzhen Stone Network Info. Ltd., 58 F.4th 785, 796 (4th Cir. 2023) (explaining that "Congress's express intent in enacting the ACPA was to curtail abusive bad faith registrations that harm commerce, business, and consumers").

87. *See* Richard A. Epstein, *The Allocation of Commons: Parking on Public Roads*, 31 J. L. STUD. S515, S515 (2002).

## B. The Economics of Accession

In modern systems of property, rights are, in the main, freely transferrable. If someone cannot effectively use a resource, it can be sold or given to another. Thus, if transfer costs were zero, then the person who can make the best use of the resource would become the owner.[89] All things equal, it would be best if rights were assigned to the person who could exploit the resource best, since society would save on the costs of the transfers. But note that Parent's method for exploiting the resource (here, the trade secrets) is identical to Orphan's. They both would use the same computer code and marketing communication channels. While spectacular and novel, it makes little difference whether an autonomous AI introduces the additional complication that one of the competitors asserting ownership (here, Orphan, Inc.) to some extent embodies the method (the code). The ability of an autonomous AI to exploit a resource is essentially identical to that of its creator. Possession is generally synonymous with ability.[90]

It may be tempting then, to assign ownership to Orphan, or at least be indifferent to Parent or Orphan as an owner, since both can exploit the resource with equal skill and maximize social welfare.[91] But this would be a mistake. Accession is socially valuable because it identifies a unique owner of a particular resource—and does so at a modest informational burden.[92] Generally, unique and unambiguous ownership of property is desirable because it encourages efficient investment in property caretaking and improvements.[93] With this simple idea in mind, it is easy to see why ownership should be assigned to Parent. If Parent knows that it will not reap the benefits of its investment in the discovery of valuable intellectual property, it will do less to develop the capabilities of Orphan. As a result, the interfacing trade secrets or patents are more likely to remain undiscovered and unused. From an economic point of view, we arrive again to the comparatively higher importance of Parent's creation efforts.

---

88. David Hume, A Treatise of Human Nature 510 n.2 (L.A. Selby-Bigge ed. 1888) (1739) (emphasis added).

89. This is true because *A* could transfer the property to *B* in exchange for a benefit greater than *A*'s benefit from exploiting the resource. *See* Ronald Coase, *The Problem of Social Cost*, 3 J. L. & Econ. 1, 16–19 (1960).

90. Differences in capital to fund, say, electricity and other costs to run the algorithm can be set aside. Since Parent and Orphan earn the same profit from exploiting the trade secrets, then they both can borrow at identical capital costs, all other things equal.

91. I have been proceeding under the assumption that the use of the trade secrets is socially beneficial. Obviously, some autonomous AI may impose social costs. This assumption will be relaxed in Part IV.

Note that the analysis has said little about the possibility of dividing the outputs of Orphan between Parent and Orphan, Inc. This is because accession, as an all-or-nothing rule, is essential for encouraging full investment in Orphan. Even if intermediate levels of investment may be sufficient to motivate Parent to create Orphan, consider that the social benefits of full investment outweigh any gains from dividing the property. This is true because any variety in the assignment of rights simply divides the pie, while any addition in investment increases, or at least maintains, its size. In summary, the economic justification for granting ownership to Parent remains compelling, even if the law eventually recognizes non-human-controlled entities capable of asserting ownership over autonomous AI.

---

92. In other words, it is mostly easy to identify whether a resource is prominently connected to another. Merrill, *supra* note 64, at 476. Note that the informational burden of assigning ownership to Parent and Orphan is more or less identical. It is easy to identify the prominent connection of Orphan to Parent, but it is comparatively easy to see Orphan's separateness and independence. In either case, property rights can be quickly established.

93. Unclear ownership of property imposes social costs. First, market efficiency is reduced because ambiguously owned property cannot be easily bought or sold, leading to a loss of economic potential. *See* Harold Demsetz, *Toward a Theory of Property Rights*, 57 AM. ECON. REV. 347, 347 (1967). Real estate with a clouded title, for instance, may remain unused or undervalued. Unclear ownership in general encourages the development of informal or gray markets that can be less efficient than formal or regulated ones insofar as they make it more difficult to transact. Relatedly, ambiguously owned property can encourage illegal trade and facilitate fraud. Second, unclear ownership encourages idleness and underutilization of property because its potential users cannot legally access or invest in it. Land may lie fallow, for example, rather than being developed or farmed. Third, unclear ownership reduces incentives for maintenance and stewardship. Ambiguously owned property is often overused, neglected, or otherwise subject to degradation. *See* Garrett Hardin, *The Tragedy of the Commons*, 162 SCIENCE 1243, 1243 (1968) (describing the problem). Cultural artifacts or natural resources may be exploited without accountability, for instance. Without clear ownership, incentives to invest in maintaining or improving property are severely curtailed. Fourth, ambiguous ownership often stifles innovation and productivity. An inability to leverage property (e.g., land and intellectual property) as collateral for loans limits access to capital and dampens entrepreneurship, hampering economic mobility. Fifth, unclear ownership can exacerbate inequality and social tensions. Ambiguous ownership often disproportionately affects marginalized groups, who may lose access to resources or face displacement. For example, indigenous communities may lose land due to unclear or undocumented ownership. Further, competing claims can lead to conflict, instability, and reduced societal cohesion. These reasons explain why society implements rules for assigning ownership.

## IV. Untraceable Ownership

We now turn to the question of what to do when autonomous AI cannot be traced to a creator. As noted in the Introduction, several scenarios are easy to imagine. For example: (1) an owner-creator loses custody of an autonomous AI that contains no identifiable information connecting it back to the owner-creator—the way one might lose a pet without a tag; (2) an AI spawns another AI with no identifiable information connecting it to its spawning parent—similar to the Tyler Cowen scenario discussed above; or (3) an owner intentionally creates an untraceable AI, perhaps with a payment system that uses an anonymous cryptocurrency address to anonymously maintain itself.

We have already seen that unassigned ownership can encourage malinvestment, and that the justification for assigning ownership to Parent is predicated on the need to encourage Parent to create autonomous AI in the first place insofar as the AI is considered socially valuable.[94] When an AI is untraceable, however, either due to carelessness or deliberate intention, assignment of rights to Parent alone is no longer welfare-enhancing in every instance. Consider the case where Parent loses track of the autonomous AI or its spawns. While this loss can be attributed to insufficient investment in monitoring, accidental misplacement does not suppress an incentive to create.[95] Parent proceeds with the initial investment as if Orphan would remain in its custody. Assigning ownership cannot be predicated upon encouraging the creation of socially valuable autonomous AI. Ownership rules must be supported by other reasons.

Cases of designed un-traceability, by contrast, can indicate deliberate misconduct. A creator may intentionally hide its prominent connection to an autonomous AI in order avoid liability for the AI's accidents or evade taxes.[96] An appropriate policy response, therefore, is to assign rights to a party capable of bearing the associated liabilities and tax obligations. In both scenarios—where un-traceability is due to carelessness or intention—assignment of property rights with a first possession rule can have a socially beneficial effect.

---

94. *See supra* Part III.
95. A careless loss does not deter the creation of what was lost because the carelessness was unintended and unexpected. For example, a rational creator may assign a probability of 10% to losing the autonomous AI, which lowers its expected value from 100 to 90. If cost of creating the AI is 91, then the investment will not be made. But if the creator assigns a zero probability of losing the AI, then its value is 100 and the investment takes place. By acknowledging that misplacement is "accidental," I am assuming that the creator assigns a probability of zero, or near-zero, to losing the AI.

It may be remarked (again) that the autonomous AI can be given legal personality and the ability to own property. It could then face liability for accidents and pay taxes itself. As stressed earlier, law requires natural persons to sit at the top of chains of ownership.[97] It is true that this rule could be set aside, but note that the problem of un-traceability would not be solved. The autonomous AI could become skilled at hiding itself, for example, by creating un-traceable autonomous AIs just like Parent. It could use these spawns to carry out work that, by all appearances, is carried out independently, outside of its control. Liability can be evaded. Taxes can be evaded just as well. The autonomous AI could create and program un-traceable spawns to collect payments for their services and then secretly transfer those payments to the Parent AI. In short, the autonomous AI could intentionally design un-traceable spawns in the same manner as the human-controlled Parent. The solution, as before, is to assign ownership to a traceable owner that can assume liability and pay the taxes.

## A. The Laws of First Possession

If accession works like a magnet, pulling in nearby property that shares a strong connection, then first possession functions more like a race, in which ownership is assigned to the one who takes control of the property first.[98] Prizes exist in an open-access commons,[99] as did tracts of land in the Oklahoma territory. At the sound of a gun, Sooners galloped off to claim one first.[100] There is competition. Wild animals, too, are typically awarded to whomever captures them first.[101] If Crockett brings an elk under his control before Houston, then the elk is Crockett's. There is a competing principle, *ratione soli*, which awards an animal to the owner of the land on which the animal is captured. This accession-based principle can govern even if the land is unenclosed.[102] And certain animals, such as bees and other animals that build homes in fixed locations—such as beavers, muskrats, and shellfish—are likely to be assigned to the owner of the land.[103] Capture of autonomous AI may be awarded to the first possessor, but these exceptions suggest that the principles of accession could justify ownership in some circumstances.

---

96. Liability can be avoided, for example, by secretly programming the "unowned" AI to carry out work for Parent. The AI, severed from Parent by all appearances, would look like an independent contractor outside of Parent's control. Any accidents caused by the AI's actions would less likely be the responsibility of the Parent. Tax evasion can be accomplished, for instance, by programming the AI to collect payments for its services and secretly transfer them to Parent.

97. *See supra* note 51.

98. This distinction is made in Merrill, *supra* note 64, at 463.

99. Sometimes the term "commons" denotes a physical (or virtual) space restricted to members of a specific community; access is controlled through social norms or law. The term "open access" can be used to denote no restrictions. *See id.* at 462 n.2.

100. *Id.*

For example, rights could be assigned to the owner of the servers where an un-traceable autonomous AI is located. If the AI is restricted to a single server or a small group of physical servers used exclusively by a person or entity, autonomous AI is more analogous to an animal captured on enclosed land or one that builds a home at a fixed location. In this scenario, ownership rights could potentially be assigned to the entity that is using the servers or other infrastructure which is hosting the AI. Many autonomous AIs, however, could be hosted on cloud servers or the Internet generally, which resembles unenclosed land or a home without a fixed location.[104] If the AI is located on this network, it could be considered unconfined, or at least minimally confined. In this case, the more fitting analogy may be to a freely roaming animal, in which rights are assigned to the initial controller.

A similar rationale distinguishes the treatment of lost and mislaid property. If Samantha finds a twenty-dollar bill lying on the floor of a restaurant, it is considered lost property and belongs to her as a first-finder—unless she is trespassing or the bill is embedded in the premises.[105] However, if the twenty-dollar bill is inside a wallet, it is typically classified as mislaid property.[106] In this case, the restaurant owner will hold it in custody for some time. If unclaimed after a reasonable period and other requirements are satisfied, ownership may transfer to the restaurant owner.[107] Thus, ownership of lost property is determined by first possession, but abandoned property is assigned through accession.

---

101. There are exceptions, however. A competing principle, *ratione soli*, awards captured wild animals to the owners of land. Honeybees are assigned to landowners. *See* State v. Repp, 73 N.W. 829, 829 (Iowa 1898). And often, animals that build homes in fixed locations—such as beavers and muskrats—are awarded on the basis of accession. *See* Merrill, *supra* note 64, at 470.

102. *See* Dale D. Goble & Eric T. Freyfogle, Wildlife Law 133–140 (2002).

103. *See* Merrill, *supra* note 64, at 470.

104. The infrastructure, for instance, could be treated as a hybrid commons. *See generally* Lawrence Lessig, The Future of Ideas: The Fate of the Commons in a Connected World (2001) (noting that the Internet originally functioned as a commons but has increasingly been privatized, and arguing for a balance between open-access and proprietary control to maintain innovation); Elinor Ostrom, Governing the Commons: The Evolution of Institutions for Collective Action (1990) (documenting hybrid systems for governing access); Michael J. Madison et al., *Constructing Commons in the Cultural Environment*, 95 Cornell L. Rev. 657 (2010) (documenting how intellectual property laws and private governance impact the commons model in online spaces).

Thus, the question arises whether un-traceable AI that is misplaced due to carelessness can be better characterized as property that is lost or mislaid. The fact that such AI is un-traceable implies that its creator cannot easily return to collect it. A reveler might remember where she left her wallet the night before and easily return the next day to the nightclub to retrieve it. This is not the case for un-traceable AI. By virtue of its un-traceability, it is truly lost insofar as its initial owner cannot trace its location. Of course, the difficulty of tracing can be a matter of degree, but even wild animals that have been captured and brought under control remain the property of the original possessor only if they are recaptured within a short period.[108] If the animal remains at large for too long, it is considered wild once again, and ownership is lost. It seems likely then, that a difficult-to-trace AI that takes time to recover will probably be understood as reentering the commons. These analogies may aid in determining rights assignment, but an examination of welfare maximization, as explored earlier with accession, can provide further insight.

---

105. *See, e.g.*, Favorite v. Miller, 407 A.2d 974, 978 (Conn. 1978) (denying a claim to property because the property was embedded in someone else's land and the finder obtained it by trespassing); Barker v. Bates, 30 Mass. 255, 261 (1832) (granting a claim for damages of a landowner for the value of timber that washed up from the sea onto his land and was taken by trespassers).

106. *See, e.g.*, Benjamin v. Lindner Aviation, Inc. 534 N.W. 2d 400, 408 (Iowa. 1995) (concluding that the finder of money concealed within the wings of an airplane did not have a right to the money because it was classified as mislaid property based on the circumstances of its concealment and the location where it was found). In some cases, ownership can pass to local authorities instead of the restaurant. *See* Merrill, *supra* note 64, at 471.

107. Law generally requires evidence of intent to abandon before passing ownership to the restaurant owner. The condition of the property, efforts to locate the owner, and reporting the loss to the state can all be of importance. Moreover, common law rules can be overlain with statutory rules that govern specific requirements related to contacting the authorities or how long the restaurant owner must wait for the original owner to reclaim the property. The point is simply that once property is deemed abandoned, ownership can be assigned by accession.

108. *See* Mullett v. Bradley, 53 N.Y.S. 781, 783 (N.Y. Sup. Ct. 1898).

### B. The Economics of First Possession

As noted earlier, ownership is valuable because it provides incentives for investment in useful labor and property improvements, enhances market function, and resolves coordination problems.[109] First possession rules are valuable because they assign ownership in a clear, simple way.[110] For instance, a finders-keeper or salvage rule prevents resources from being wasted by ensuring that they are recovered and put back into circulation. In addition, these rules lower the costs of resolving disputes by providing a predictable framework for determining ownership.

The main drawback of the first possession principle is that it can result in duplicative effort when more than one person competes to claim the same property.[111] If A and B gallop off to claim a tract of land, and B arrives first, then A's effort to claim the land was made in vain and was socially wasteful. This problem is mitigated if A and B possess complete information about one another's talents. If A knows that B is faster, then A will not wastefully compete.[112] Even in the presence of duplicative effort, first possession is efficient insofar as the social benefits of bringing property under the possessor's control are greater than the costs of competition.

### V. The Deployment of Ownership Rules

It is now time to consider how accession and first possession rules could be operationalized to encourage ownership of untraceable autonomous AI that has been carelessly misplaced or intentionally created. A preliminary assumption is that the autonomous AI exists as software or code. This code can be present on local physical servers or hosted on a cloud such as Amazon Web Servers, Google

---

109. *See supra* note 93.
110. *See* Richard A. Epstein, *Possession as the Root of Title*, 13 Ga. L. Rev. 1221, 1234 (1979); Carol Rose, *Possession as the Ownership of Property*, 52 U. Chi. L. Rev. 73, 73 (1985).
111. *See* Merrill, *supra* note 64 at 482–83.
112. *See* Christopher Harris & John Vickers, *Perfect Equilibrium in a Model of a Race*, 52 Rev. Econ. Stud. 193, 194 (1985) ("[I]f one player is far enough ahead of the other . . . then the latter gives up completely, leaving the former to move to the finishing line at his own pace."). Unfortunately, there is a worst-case equilibrium in which potential claimants can acquire the technology guarantee that they are able to capture the resource first. If the technology is available to both claimants at equal cost, then they can reach the resource at the same time. However, this is unlikely insofar as claimants possess some advantage prior to investing in the technology or if there is some random variability in opportunities. *See* Dean Lueck, *First Possession*, *in* The New Palgrave Dictionary of Economics and the Law 133, 133–42 (1998).

Cloud, or Microsoft's Azure.[113] Consider the case of local physical hosting. Suppose an employee of Startup creates an AI, which is hosted on the employer's server. The employee leaves the firm. After some time, the AI spawns an autonomous AI that is hosted on the firm's server. Because it is untraceable, no one knows its provenance. However, because the autonomous AI sits on the firm's servers, and no one outside the firm can access them, it can be inferred that it originated within the firm, especially if it can be shown that the servers have strict requirements for connecting to the outside world. This is probably a good case for an accession rule that assigns ownership to the firm.

Consider the scenario in which the autonomous AI is programmed to intentionally migrate from a private server to a cloud. This can be accomplished through automated uploads, deployment scripts, and even backup services. Now the untraceable code is located outside the firm. Certainly, a better case can be made for first possession, since the firm's prominent connection to the code has been lost. Perhaps the firm could come forward with an older version of the code that it found on its own server and assert a connection, but the point is that when a connection is severed, the rationale for assigning rights based on accession is weakened.

How should we interpret the fact that the code resides on a privately owned cloud? Suppose it moves to Amazon Web Services. In a way, it becomes "enclosed" within Amazon's cloud infrastructure. Yet Amazon played no active role in attracting the autonomous AI beyond simply offering cloud services as part of its business model. And even though the autonomous AI is located on Amazon's cloud, it is not under Amazon's control. Amazon possesses no more incentive than you or me to invest in improving the code; market the code's outputs more widely; or otherwise enhance the code's social value. It seems more fitting then, that a rule of first possession should apply. This rule encourages competition to "capture" the untraceable AI and assign ownership to a controller who can increase its social value as quickly as possible. It may be remarked that ownership should be granted to the party best positioned to maximize the social value of the autonomous AI—such as by ensuring its outputs are widely and effectively distributed. However, any system of ownership assignment naturally promotes this outcome, as the custodian who is granted ownership can subsequently transfer the property to the user who values it most. Welfare is maximized by assigning ownership as quickly as possible. This is accomplished with first possession.

---

113. There can be layers to hosting. For example, the autonomous AI may be located on GitHub, but GitHub primarily relies on Microsoft Azure for hosting. The same configuration can arise under local control. GitHub Enterprise Server allows business to host an autonomous AI on its own servers or cloud environments.

At least one transfer is likely if the code is encrypted, and the key remains unavailable. Decryption requires specialized expertise and significant computational resources.[114] A decryption specialist is unlikely to fully capitalize on every autonomous AI it acquires. Instead, it will decrypt and take access of the code only when the potential benefits—such as selling it to a third party capable of utilizing the AI—exceed the costs of decryption. Note that taking custody need not be the only way to determine first possession. As implicated in *Pierson v. Post*,[115] hot pursuit or some other norm for establishing custody could be enough.[116] Decrypted access is a useful marker because it is easy to verify. One either has control of the code or not.

When the costs of decryption and capture outweigh the benefits of ownership, the government or private charities can stimulate competition through subsidies. Notably, asserting default government ownership offers no advantage, as control remains unattainable until the AI is secured. In fact, such a claim could hinder the process by eliminating competitive incentives—without rivals, the government would face little urgency to act swiftly in bringing the AI under control. There is a special case, raised in the Introduction, in which the highest-valued user of the AI is the government. Today, this often occurs when there are strong positive externalities, as is the case with national defense. Free automated tax services for people with incomes under $75,000 could be an example of this. If the government's interfacing costs between a private provider and the IRS are high, then it may be efficient for the government to simply offer the service. This is more likely to be the case when the costs of offering the service are low, as is the case with an automated tool. This special case offers an avenue for the government to assert ownership.

Finally, consider the case of an autonomous and untraceable AI carrying out malicious activities such as automated hacking, deepfake impersonation, phishing, market manipulation, and the automated trafficking of illegal goods. If the malicious AI is hosted on a private server, control (including ownership) can be assigned to the server's owner based on a prominent connection. Even if the AI is encrypted, the owner can power off the server, remove drives, or restrict access. Insurance could cover associated costs. If the AI is hosted on the cloud, it makes less sense to attribute control (and assign ownership) unless the host can easily quarantine or eliminate it.[117] But imagine a scenario in which the malicious AI cannot be easily located. For instance, the malicious AI could blend in with normal traffic and activity by mimicking legitimate users, using common protocols, or disguising itself as a background process. It could adapt and modify itself to avoid discovery by using polymorphic code to avoid signature detection, only run under low-risk conditions, or shut down or behave benignly when in high-risk conditions. In general, it could become skilled at hiding. Governments or private charities could offer bounties for taking custody of the malicious code. In principle, a bounty system would serve as first possession-style rule in which the person who eliminates or takes custody first receives the reward. Such a system is efficient insofar as the benefits of capture exceed the cost of the bounty and its administration. Note that the government itself can participate in the competition. If it captures the AI first, then the bounty simply goes unpaid.

---

114. Cloud service agreements and legislation prohibit the hacking of access credentials. *See, e.g.*, 18 U.S.C. § 1030 (providing criminal penalties for accessing a computer without authorization to obtain certain restricted or protected information). But these can be changed to permit capture of untraceable AI.

115. Pierson v. Post, 3 Cai. R. 175 (1805).

116. *Cf. id.* at 178–80.

## VI. Conclusion

The rise of autonomous AI presents fundamental challenges to traditional conceptions of ownership. As AI systems become increasingly self-sufficient in production, distribution, and even self-replication, the legal frameworks that assign property rights must evolve to accommodate new realities. The conventional mechanisms of accession and first possession provide useful pathways for determining ownership when AI can be traced to an originator. However, when AI becomes untraceable—whether through carelessness, deliberate obfuscation, or emergent behavior—these principles require modification to prevent inefficiencies and regulatory arbitrage.

For traceable AI, ownership can be efficiently assigned based on accession, ensuring that those who create and deploy autonomous systems retain control over their outputs and liabilities. This encourages investment in AI development while maintaining accountability. However, for untraceable AI, particularly AI that has either lost its connection to an identifiable owner or has been designed to evade attribution, a first possession rule emerges as a viable alternative. This rule ensures that AI remains within the realm of human governance, as those who capture and control such entities are incentivized

---

117. Generally, liability is assigned to the host if it knowingly facilitates illegal activity. *See, e.g.*, 18 U.S.C. § 1030(a) (prohibiting knowingly hosting an AI that facilitates hacking, unauthorized access, or cyberattacks that fall under "aiding and abetting" computer crimes).

to either integrate them productively or bring them under regulatory compliance.

Where ownership remains ambiguous or contested, alternative approaches—such as bounty systems for capturing rogue AI, or private incentives to encourage competition in AI capture—may become necessary. Importantly, default government ownership is ineffective, as the government does not gain control over an unclaimed AI until it is secured. Without competition, government claims would only slow the race to capture rogue AI, reducing incentives for rapid intervention. Instead, targeted subsidies or market-driven incentives can encourage competitive efforts to bring untraceable AI under control, aligning private and public interests in enforcement.

Ultimately, whether AI-driven automation leads to the dilution of ownership or its strategic redeployment, the law must adapt to balance innovation with accountability. Ensuring that AI remains within a framework of assignable rights—whether through accession, first possession, or alternative models—will be crucial in shaping a future where autonomy does not mean anarchy, and where technological advancement remains aligned with societal order.